\documentclass[aip,jcp,amsmath,amssymb,reprint,nofootinbib,floatfix,longbibliography]{revtex4-2}

\usepackage[utf8]{inputenc}
\usepackage[T1]{fontenc}
\usepackage{lmodern}
\usepackage{mathtools}
\usepackage{bm}
\usepackage{graphicx}
\usepackage{xcolor}
\usepackage{tikz}
\usepackage{pgfplots}
\pgfplotsset{compat=1.17}
\usepackage[colorlinks=true,linkcolor=black,citecolor=black,urlcolor=black]{hyperref}
\usepackage{orcidlink}
\usepackage{placeins}
\usepackage{float}

\definecolor{approvedblue}{rgb}{0,0,0}
\definecolor{revgreen}{rgb}{0,0,0}
\definecolor{issuered}{rgb}{0,0,0}
\definecolor{fixpink}{rgb}{0,0,0}
\newcommand{\new}[1]{{\color{revgreen}#1}}
\newcommand{\upd}[1]{{\color{approvedblue}#1}}

\definecolor{refblue}{rgb}{0,0,0}
\definecolor{refred}{rgb}{0,0,0}
\definecolor{refgreen}{rgb}{0,0,0}
\newcommand{\rev}[1]{{\color{refblue}#1}}

\definecolor{audblue}{rgb}{0,0,0}
\definecolor{audred}{rgb}{0,0,0}
\definecolor{audpink}{rgb}{0,0,0}
\newcommand{\aud}[1]{{\color{audblue}#1}}

\definecolor{pitblue}{rgb}{0,0,0}
\definecolor{pitred}{rgb}{0,0,0}
\definecolor{pitgreen}{rgb}{0,0,0}
\newcommand{\pit}[1]{{\color{pitblue}#1}}

\definecolor{milblue}{rgb}{0,0,0}
\definecolor{milyellow}{rgb}{0.72,0.53,0}
\definecolor{milred}{rgb}{0.85,0,0}
\definecolor{milgreen}{rgb}{0,0.50,0.16}
\newcommand{\mil}[1]{{\color{milblue}#1}}

\definecolor{harblue}{rgb}{0,0,0}
\definecolor{harred}{rgb}{0.85,0,0}
\definecolor{hargreen}{rgb}{0,0.50,0.16}
\newcommand{\har}[1]{{\color{harblue}#1}}

\makeatletter
\g@addto@macro\normalsize{%
  \setlength\abovedisplayskip{4pt plus 1pt minus 2pt}%
  \setlength\belowdisplayskip{4pt plus 1pt minus 2pt}%
  \setlength\abovedisplayshortskip{2pt plus 1pt}%
  \setlength\belowdisplayshortskip{2pt plus 1pt}%
}
\makeatother

\makeatletter
\def\rp@gnobreaktrue{\global\let\if@nobreak\iftrue}
\def\rp@gnobreakfalse{\global\let\if@nobreak\iffalse}
\AtBeginDocument{%
  \def\@afterheading{%
    \rp@gnobreaktrue
    \global\everypar{%
      \if@nobreak
        \rp@gnobreakfalse
        \clubpenalty\@M
        \if@afterindent\else{\setbox\z@\lastbox}\fi
      \else
        \clubpenalty\@clubpenalty
        \everypar{}%
      \fi}}%
}
\makeatother

\newcommand{\phigold}{\varphi}                  
\newcommand{\Jcost}{J}                          
\newcommand{\Lrod}{\ell}                        
\newcommand{\Drod}{D}                           
\newcommand{\nhat}{\hat{\mathbf{n}}}            
\newcommand{\Rsph}{R_{\mathrm{s}}}              
\newcommand{\eyhat}{\hat{\mathbf{e}}_\theta}    
\newcommand{\ephihat}{\hat{\mathbf{e}}_\phi}    
\newcommand{\erhat}{\hat{\mathbf{e}}_r}         
\newcommand{\tilt}{\omega}                       

\newlength{\omegaht}
\newlength{\panellblshift}
\newlength{\panelnoteshift}
\begin{document}

\title{Tilt control of coverage heterogeneity for hard spherocylinders
locked on a sphere}

\author{Jonathan Washburn\,\orcidlink{0009-0001-8868-7497}}
\affiliation{Recognition Physics Institute, Austin, TX, USA}

\author{Hartmut L\"owen\,\orcidlink{0000-0001-5376-8062}}
\affiliation{Institut f\"ur Theoretische Physik II: Weiche Materie,
  \mbox{Heinrich-Heine-Universit\"at D\"usseldorf, Germany}}

\author{Elshad Allahyarov\,\orcidlink{0000-0001-7212-4713}}
\email[Author to whom correspondence should be addressed: ]{elshad.allakhyarov@case.edu}
\affiliation{Recognition Physics Institute, Austin, TX, USA}
\affiliation{Institut f\"ur Theoretische Physik II: Weiche Materie,
  \mbox{Heinrich-Heine-Universit\"at D\"usseldorf, Germany}}
\affiliation{Department of Physics, Case Western Reserve University,
  \mbox{Cleveland, OH 44106, USA}}
\affiliation{Theoretical Department, Joint Institute for High Temperatures,
  RAS, \mbox{Moscow 125412, Russia}}

\begin{abstract}
\har{We study hard spherocylinders on a sphere with axes rigidly locked to
a tangential director field at fixed angle $\tilt$ to the meridian,
necessarily singular at the poles.  Three lengths set the problem: the rod
length $\Lrod$, the diameter $\Drod$, and the host radius $\Rsph$.
Monte Carlo simulations across fifteen geometries and four coverages give
two main results for the polar marginal, the azimuthal average of
rod-center density.  First, the tilt is a continuous handle on the width of
the depleted region that packing induces around each singularity.  Under
meridian locking it is set principally by the rod length; turning
the director toward the latitude contracts it substantially, the
contraction being spent well before latitude locking.  Second, how uniform the
polar marginal can be made is limited by geometry, not tilt.  The smallest
variance on the sampled grid follows a power law in
$\Lrod^{2}/(\Rsph\Drod)$, which measures how far a straight rod's ends
stand off the curved surface in rod diameters, with an effective exponent
between $1.1$ and $1.3$.  Long rods on small hosts cannot be made uniform
at any sampled tilt; the remedy is geometric, not orientational.  The polar
marginal is
equator-heavy almost everywhere, inverting only at high coverage and tilt;
meridian locking is the least uniform choice, and the variance-minimizing
tilt usually lies in a sampled band from $31.7^{\circ}$ to $55^{\circ}$.  The
golden-ratio tilt $\arctan(1/\phigold)$ is one of those angles: a
benchmark, not one the model selects.  Both results describe the infinitely
locked athermal ensemble.}
\end{abstract}

\maketitle

\section{Introduction}
\label{sec:intro}

How rod-like particles cover a curved surface is a basic problem in
soft matter, with direct bearing on the design of uniform colloidal
coatings.  On a sphere the problem is constrained by topology: a
direction field everywhere tangent to a sphere cannot be continuous at
every point~\cite{Milnor1978Hairy,Nelson2002,Bowick2009}, so any
orientational order on a spherical host must contain singularities.
\mil{The Poincar\'e--Hopf theorem fixes the total index of such a field at
$+2$, but neither where the singularities sit nor what the particles do
near them: their placement follows from the field one imposes, and here
both sit at the poles because the locked field puts them there.}
What topology does not prescribe is how the particle density
distributes around those singularities.

The established model for this question is the hard spherocylinder, the
minimal model of lyotropic liquid crystals, whose ordering is driven by
entropy
alone~\cite{Onsager1949,Frenkel1988,Bolhuis1997,GrafLowen1999,deGennesProst}.
On spherical hosts, hard-rod fluids have been studied in spherical
cavities~\cite{Cleaver2000,Trukhina2008} and spherical
monolayers~\cite{Smallenburg2016,Allahyarov2017,Allahyarov2018,Mandal2026,Mandal2025Melting,Mandal2025Freezing},
and related tangent-order constraints appear in liquid-crystal
shells~\cite{FernandezNieves2007,LopezLeon2011,Liang2011,Sharma2024,Jull2024,Sato2024}
and rod membranes~\cite{Lagerwall2012}.
Recent theory has treated
active and chiral curved
smectics~\cite{Nestler2024,Caprini2024Vortices,Caprini2025Wrap,Musacchio2026},
biaxial order~\cite{ElMoumane2024,ElMoumane2025}, and anisotropic
Pickering coatings of
droplets~\cite{Shrivastava2024,Pradhan2025,Witt2024,AllahyarovLowen2024}.
A controlled variant of the problem locks the rod orientations
externally.  Particle-resolved Monte Carlo work with such locked
orientations has so far fixed the rod axes to the meridian or latitude
fields only and mapped smectic and segregation textures at those two
extreme tilts~\cite{Allahyarov2017,Allahyarov2018}. A companion
Communication~\cite{AllahyarovSmC} used the same locked field at
intermediate angles to study \har{the curvature-induced smectic-C order
that appears at the highest coverage}.
\har{What the present paper adds to that Communication should be said at
the outset.  The Communication asked what phase the rods form, and answered
it at one coverage, $\eta=0.75$, where the tilted locking produces
smectic-C layering.  The present work asks a different question of the same
locked field: not what phase appears, but where the particles sit.  We map
the azimuthally averaged density continuously in the locked angle and over
three further coverages down to $\eta=0.2$, and we use that map to separate
two effects that the single-coverage study could not distinguish, namely a
defect width that the anchoring angle controls and a residual
nonuniformity that the anchoring angle cannot remove.  The $\eta=0.75$
geometries are the only ones shared with the Communication, and they are
re-analyzed here through observables it did not use.}
\rev{\har{The present question is also distinct from}
recent particle-resolved work on freely orientable soft rods on a
sphere~\cite{Mandal2026}, where curvature, aspect ratio and density
control the phase behavior and the defect structure of a relaxing
director.  Here the director is prescribed, so the question is not which
texture the rods select but how the density responds to a texture imposed
on them.}

The unresolved gap is the density response between the two extreme tilts: how
the density averaged over a circle of constant latitude \aud{(the polar
marginal)} varies with the locked angle across sphere radii, rod
lengths, and projected-area fractions. 
This gap matters because the polar marginal is the \pit{simplest
descriptor of where the particles sit that a coating design would have to
control}.  Without mapping it, one cannot say whether the
extreme-tilt textures are typical or exceptional, whether intermediate
locked angles make the coverage more or less uniform, whether the
sign of the equator-versus-pole contrast can change under particular
geometric conditions, or what \rev{limits} how uniform the
\rev{polar marginal} can be made.
\new{Two questions in that list are the practically decisive ones.  The
first is how far the imposed tilt can push back the depleted region
\mil{induced by packing
around a topologically required singularity}, since that region is the defect
a coating has to live with.  The second is whether the residual
nonuniformity that survives the best \rev{sampled} choice of tilt is a
property of the tilt or of the geometry, since only the latter would place
a limit that \rev{no sampled locking angle removes}.}

\upd{We map the dependence of the polar marginal on the locked angle by
Monte Carlo simulation, locking hard spherocylinders on a sphere to the
tangential field that makes a fixed angle $\tilt$ with the local
meridian.} 
From each
simulation we compute the area-weighted variance
$v$ of the azimuthally averaged density $\hat\rho(\theta)$, the
equator-minus-pole contrast $\Delta_{\rm eq}$ \aud{and the polar
first-crossing angle $\theta_{1/2}$, all three defined in
Sec.~\ref{sec:theory}}.

\har{Only three lengths enter the problem, and it is worth naming them
before the results because every result below is expressed through them.
They are the rod length $\Lrod$, which is the length the rod brings to a
singularity; the rod diameter $\Drod$, which sets the thickness of the
monolayer and so the scale on which a geometric mismatch can be absorbed;
and the host radius $\Rsph$, which sets the curvature the rod has to sit
on.  The combination $\Lrod^{2}/(\Rsph\Drod)$ that appears below joins all
three, and Sec.~\ref{sec:scaling} shows why that particular combination is
the one the residual nonuniformity follows.}

\new{We obtain two main results.  The first is that the locked tilt acts
as a continuous control on the width of the depleted region around each
singularity.  Under meridian locking that width is set \har{principally} by the rod
length rather than by the host radius or the coverage, and turning the
director toward the latitude direction contracts it by nearly an order of
magnitude before the contraction saturates \har{at the resolution floor of
the binned estimator, which makes that factor a lower bound}.  The
second is that the \rev{polar-marginal} uniformity attainable by choosing
\rev{among the sampled tilts} is
limited by the geometry and not by the tilt: the lowest variance sampled
in each geometry follows an empirical scaling in
$\Lrod^{2}/(\Rsph\Drod)$, the out-of-surface mismatch of a straight rod
expressed in rod diameters, with \har{an effective exponent} between $1.1$
and $1.3$ at
\mil{each of the four densities sampled, most securely at the three lower
ones}.
Alongside these, the polar marginal is
usually denser near the equator than near the poles (equator-heavy),
meridian locking carries the larger extreme-tilt variance, \aud{at low and
intermediate coverage the entire tilt dependence of the uniformity is
carried by the ratio of the equatorial to the polar mean alone,} and rare sign
inversions of the equator--pole contrast appear at high density and high
tilt.  A golden\pit{-ratio} reference \pit{tilt, one of the sampled
angles,} lies \upd{at the lower edge of} the
low-variance band without minimizing the variance there.}

\har{The paper is organized as follows.  Section~\ref{sec:theory} sets out
the theoretical background: the locked director field, the three
observables computed from the polar marginal, the ratio-symmetric
imbalance measure, and the golden-ratio reference tilt.
Section~\ref{sec:model} describes the Monte Carlo protocol together with
the geometries, coverages, and locked angles sampled.
Section~\ref{sec:equator} reports the results.  It begins with the
prevailing equator-heavy coverage and the state of the two singularities
in Sec.~\ref{sec:profiles} and with the dependence of the coverage
uniformity on the locked angle in Sec.~\ref{sec:tilt};
Secs.~\ref{sec:jcost} and~\ref{sec:goldenref} then ask what the imbalance
measure and the golden-ratio reference add to that dependence.  The two
main results follow.  Section~\ref{sec:core} shows that the locked tilt
controls the width of the depleted region around each singularity, and
Sec.~\ref{sec:scaling} that the uniformity attainable among the sampled
tilts is limited instead by the geometric combination
$\Lrod^{2}/(\Rsph\Drod)$.  Section~\ref{sec:discussion} reads the two
results together, sets out the limitations of the externally locked model,
and concludes.}

\section{Theoretical background}
\label{sec:theory}

\begin{figure}[!t]
\centering
\newcommand{\drawSphereTilt}[3]{%
    \begin{tikzpicture}[scale=0.62, yscale=0.95]
    \pgfmathsetmacro{\Rdraw}{1.7}
    \pgfmathsetmacro{\rodL}{0.45}
    \pgfmathsetmacro{\rodD}{0.10}
    \draw[thick] (0,0) circle (\Rdraw);
    \draw[gray,thin]
      (0,0) ellipse ({\Rdraw} and {0.18*\Rdraw});
    \foreach \aphi in {-60,-30,30,60}{
      \pgfmathsetmacro{\xrad}{\Rdraw*abs(cos(\aphi))}
      \draw[gray,thin,dashed]
        (0,\Rdraw) arc[start angle=90, end angle=-90,
                       x radius=\xrad, y radius=\Rdraw];
    }
    \draw[gray,thin,dashed] (0,\Rdraw) -- (0,-\Rdraw);
    \foreach \alat in {30,60,120,150}{
      \pgfmathsetmacro{\zlat}{\Rdraw*cos(\alat)}
      \pgfmathsetmacro{\rlat}{\Rdraw*sin(\alat)}
      \draw[gray,thin,dashed,opacity=0.55]
        (0,\zlat) ellipse ({\rlat} and {0.18*\rlat});
    }
    \fill (0, \Rdraw) circle (1.5pt) node[above]{\scriptsize N};
    \fill (0,-\Rdraw) circle (1.5pt) node[below]{\scriptsize S};
    \foreach \theta in {25,55,85,115,145}{
      \foreach \phi in {-40,-20,0,20,40}{
        \pgfmathsetmacro{\xR}{\Rdraw*sin(\theta)*sin(\phi)}
        \pgfmathsetmacro{\yR}{\Rdraw*cos(\theta)}
        \pgfmathsetmacro{\mx}{cos(\theta)*sin(\phi)}
        \pgfmathsetmacro{\my}{-sin(\theta)}
        \pgfmathsetmacro{\mn}{sqrt(\mx*\mx+\my*\my)+0.0001}
        \pgfmathsetmacro{\meridAng}{atan2(\my/\mn, \mx/\mn)}
        \pgfmathsetmacro{\rodAng}{\meridAng + #1}
        \draw[fill=red!75!black, draw=red!90!black, line width=0.2pt,
              rotate around={\rodAng:(\xR,\yR)}]
          ({\xR-\rodL/2},{\yR-\rodD/2}) rectangle
          ({\xR+\rodL/2},{\yR+\rodD/2});
      }
    }
    \node[below=\panellblshift] at (0,-\Rdraw) {\scriptsize #2};
    \node[font=\tiny, text width=2.6cm, align=center,
          below=\panelnoteshift] at (0,-\Rdraw) {#3};
  \end{tikzpicture}%
}
\scalebox{1}[0.95]{\drawSphereTilt{0}{$\tilt = 0$ (meridian)}{}}\hfill
\scalebox{1}[0.95]{\drawSphereTilt{45}{$\tilt = 45^{\circ}$ (tilted)}{}}\hfill
\scalebox{1}[0.95]{\drawSphereTilt{90}{$\tilt = 90^{\circ}$ (latitude)}{}}
\caption{Schematic side views of hard spherocylinders tangentially
anchored on the host sphere of radius $\Rsph$, for three reference
tilts of the locked director field
$\nhat$ 
[Eq.~\eqref{eq:tiltfield}].  Dashed gray curves mark meridians and
parallels (orientation only), and red rectangles are front-facing rods.
\har{The labels N and S in each panel mark the north and south poles of the
host sphere, that is the two points $\theta=0$ and $\theta=180^{\circ}$ at
which the tangential director field is singular.}
Left: meridian locking \upd{at $\tilt=0$}.  Center: intermediate locking
\upd{at $\tilt=45^{\circ}$, labelled tilted in the panel}.
Right: latitude locking \upd{at $\tilt=90^{\circ}$}.
\pit{The panels are drawings of the imposed director geometry, not
simulation output, and no density information is shown in them.}
The drawing 
is not a geometrically
exact projection of the three-dimensional tangent field.}
\label{fig:sphere_geom}
\end{figure}
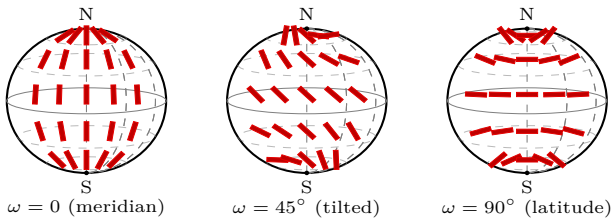

The system is a fluid of monodisperse hard spherocylinders (cylinder
length $\Lrod$, diameter $\Drod$) whose centers lie on a sphere of
radius $\Rsph$ and whose long axes are rigidly locked to the tangential
director field
\begin{equation}
  \nhat(\theta,\phi)
  =\cos\tilt\,\eyhat+\sin\tilt\,\ephihat ,
  \label{eq:tiltfield}
\end{equation}
where $\theta$ and $\phi$ are the colatitude and azimuth on the sphere,
$\eyhat$ and $\ephihat$ are the meridian and latitude unit
vectors, and $\tilt$ is the imposed tilt angle: $\tilt=0$ recovers
meridian locking and $\tilt=90^{\circ}$ latitude locking.  For
intermediate tilts the integral curves of the field are loxodromes, the
constant-bearing curves that cross every meridian at the same angle.
\mil{Near either pole the sphere is locally a plane, with radial distance
$r\simeq\Rsph\theta$ from the singular point, and
Eq.~\eqref{eq:tiltfield} becomes
$\nhat\simeq\cos\tilt\,\erhat+\sin\tilt\,\ephihat$ with $\erhat$ radial in
that plane.  The locked tilt therefore interpolates between the three
standard planar textures of a unit-charge defect: $\tilt=0$ is a radial
aster, intermediate tilts are spirals, and $\tilt=90^{\circ}$ a vortex.
All carry the same index, so the tilt changes the geometric realization of
the singularity and not its topological charge.}
The same locked field underlies the companion
Communication~\cite{AllahyarovSmC}, from which the panels of
Figure~\ref{fig:sphere_geom} are adapted.  Those panels illustrate the
three orientation classes (meridian, intermediate, and latitude
locking) as schematic views of the director field, not density maps. 
Every quantitative density profile reported here is obtained 
from Monte Carlo simulation.

\mil{The locked field also has a compact differential-geometric reading
that will be useful in interpreting the results.  Away from the poles the
splay and bend of Eq.~\eqref{eq:tiltfield} are
\begin{equation}
  \nabla\!\cdot\!\nhat=\frac{\cos\tilt}{\Rsph}\cot\theta,
  \qquad
  |(\nhat\!\cdot\!\nabla)\nhat|=\frac{|\sin\tilt|}{\Rsph}|\cot\theta| ,
  \label{eq:splaybend}
\end{equation}
so that
\begin{equation}
  (\nabla\!\cdot\!\nhat)^{2}+|(\nhat\!\cdot\!\nabla)\nhat|^{2}
  =\frac{\cot^{2}\theta}{\Rsph^{2}} ,
  \label{eq:distortion}
\end{equation}
independently of $\tilt$.  Turning the director therefore does not reduce
the local distortion of the field; it converts splay into bend at fixed
total.  Two consequences matter here.  The distortion still diverges at
both poles at every tilt, which is the field-level statement that the
singularity cannot be removed.  And because the two terms enter a
one-constant Frank energy with equal weight, that energy is the same at
every tilt, so a purely elastic reading of the imposed field would predict
no tilt dependence at all; whatever tilt dependence the simulations show
has to come from the packing of the rods.}

The quantity we analyze throughout is the polar marginal
$\hat\rho(\theta)$, the density averaged over azimuth at fixed
colatitude $\theta$ and normalized to unity for uniform coverage.
\pit{The estimator is a histogram of rod centers: the number of centers
whose colatitude falls in bin $i$ is divided by the surface area of that
bin and by the mean areal number density $N/(4\pi\Rsph^{2})$, so
$\hat\rho_i=1$ means the bin holds its share of the rods.  It is a
center density and not the fraction of surface area covered by rod
bodies, which for rods of finite length spanning several bins is a
different quantity.  Uniformity below always means uniformity of this
center density.}  We
compute three observables from it.
The first is the area-weighted polar-marginal variance
\begin{equation}
  v=\sum_i w_i[\hat\rho(\theta_i)-1]^2,
  \qquad
  w_i=\frac{A_i}{4\pi\Rsph^2}
  \label{eq:variance}
\end{equation}
where the sum runs over the angular bins centered at $\theta_i$, the
binned profile value is $\hat\rho_i\equiv\hat\rho(\theta_i)$, and $A_i$
is the surface area of bin $i$, so that the weights $w_i$ sum to one.
\har{The bins are equally spaced in colatitude rather than in area, so
$A_i\propto\sin\theta_i$ and a bin at the equator covers far more of the
sphere than a bin at a pole.  That inequality is what the weights $w_i$
remove: each bin enters $v$ in proportion to the area it occupies, so the
crowded polar bins carry no more influence on $v$ than their share of the
surface, and equal-angle binning biases the measure toward neither the
poles nor the equator.  What the bin width does control is resolution,
since structure finer than a bin is averaged away, and that is a separate
matter taken up with $\theta_{1/2}$ below and in
Sec.~\ref{sec:discussion}.}
\aud{We emphasize that $v$ characterizes the polar marginal only and is not
a measure of full-surface coverage:}
\rev{because the azimuthal average is taken before the variance, $v$
carries no information about structure around a circle of constant
latitude, so a coverage with longitudinal stripes or with smectic layers
running across the poles can have a flat polar marginal, and every
uniformity statement made below is to be read in this restricted sense.}
\mil{Equivalently, the average retains only the axisymmetric $m=0$ sector
of a spherical-harmonic expansion of the density, so $v$ is a
symmetry-reduced measure of uniformity and every $m\neq0$ modulation is
invisible to it by construction; a full-surface variance, or the spectrum
of coefficients $a_{\ell m}$, would lift the restriction.}
The smallest variance sampled in a
cell is denoted $v_{\min}$, which is 
\rev{the minimum over the discrete set of locked angles
  with no interpolation between the sampled angles.}

The second observable is the equator excess
\begin{equation}
  \Delta_{\rm eq}
  =\langle\hat\rho\rangle_{60^{\circ}\le\theta\le120^{\circ}}
  -\langle\hat\rho\rangle_{\theta<30^{\circ}\ {\rm or}\ \theta>150^{\circ}}
  \label{eq:deq}
\end{equation}
where $\langle\hat\rho\rangle_A$ denotes the area-weighted average of
$\hat\rho$ over the region $A$.  The equatorial belt is therefore
$60^{\circ}\le\theta\le120^{\circ}$, and the polar caps are
$\theta<30^{\circ}$ and $\theta>150^{\circ}$.  A positive
$\Delta_{\rm eq}$ means that the coverage is denser near the equator
than near the poles.

The third observable 
is the polar first-crossing angle $\theta_{1/2}$, defined as the center of the
first angular histogram bin away from a pole at which $\hat\rho$ reaches
one half of the uniform value.  \har{The subscript denotes that half-density
level and is not a pair of indices: $\theta_{1/2}$ is a single angle, one
number per panel, and not two angles.}  The corresponding arc length on the
sphere is $r_c=\Rsph\theta_{1/2}$.
\rev{The rule is applied by scanning outward from each of the two poles
in turn and averaging the two crossings, and it has two properties that
bound what it can resolve.  A crossing found in the bin that touches the
pole means only that the true crossing lies somewhere inside that bin, so
the reported value is then an upper bound fixed by the bin width rather
than a measured width.  The same applies when the pole bin is already
denser than one half of the uniform value, as happens under latitude
locking, since the scan then terminates immediately and returns the center
of that first bin.  Values of $\theta_{1/2}$ reported at large tilt are
therefore upper bounds, and the contractions quoted in
Sec.~\ref{sec:core} are correspondingly lower bounds.}

Alongside the signed equator excess we introduce a dimensionless,
ratio-symmetric measure of the equator--pole imbalance.  With the
area-weighted regional means
\begin{equation}
  \mu_{\rm eq}
  =\frac{\sum_{{\rm eq}}w_i\hat\rho_i}{\sum_{{\rm eq}}w_i},
  \
  \mu_{\rm pole}
  =\frac{\sum_{{\rm pole}}w_i\hat\rho_i}{\sum_{{\rm pole}}w_i},
  \
  x=\frac{\mu_{\rm eq}}{\mu_{\rm pole}}
  \label{eq:xratio}
\end{equation}
where the equatorial belt and polar caps are those of
Eq.~\eqref{eq:deq}, the binned profile values $\hat\rho_i$ and the area
weights $w_i$ are those of Eq.~\eqref{eq:variance}, and the same
weights enter both means.
Because both regional means are positive, $x$ is positive throughout
and $\Delta_{\rm eq}<0$ holds exactly when $x<1$.  A coverage inversion
is therefore defined once, by $\Delta_{\rm eq}<0$, and $\Delta_{\rm eq}$
is used for all sign counts reported below.

We also introduce a ratio-symmetric transform function
\begin{equation}
  \Jcost(x)=\frac12\left(x+\frac1x\right)-1,
  \label{eq:Jx}
\end{equation}
where $x$ is the regional means ratio defined in Eq.~\eqref{eq:xratio}. 
This transform is invariant under the interchange of equator and
poles, vanishes at balance $\Jcost(1)=0$, and grows with the magnitude of the
imbalance.
In the present work $\Jcost$ is used only as a diagnostic of a measured
profile; it does not predict the tilt.  The ratio $x$ becomes noisy
when $\mu_{\rm pole}$ is small, so we attach no physical threshold
to $x$ or $\Jcost(x)$ and do not use them to classify individual panels.
\new{We report instead the summary statistics of $\Jcost$ over the whole
data set in Sec.~\ref{sec:jcost}\aud{, where what the measure does and does
not resolve is established from the data}.}

As a reference we also evaluate the golden\pit{-ratio} tilt
\begin{equation}
  \tilt_\phigold=\arctan(1/\phigold),
  \qquad
  \phigold=\frac{1+\sqrt5}{2},
  \label{eq:golden}
\end{equation}
with $\phigold$ the golden ratio.  This choice is motivated, not
derived: $\phigold$ is the unique positive fixed point of
\aud{$t\mapsto1+1/t$}.  We stress that
$\tilt_\phigold$ is a benchmark tilt, not a predicted optimum; the
simulations test where the variance minimum actually lies relative to
it.  \pit{Nothing in the hard-rod model singles out this angle, no
packing argument selects $\arctan(1/\phigold)$, and it enters below only
as one of the sampled grid angles of Eq.~\eqref{eq:tiltgrid} against which
the measured behavior is read.  It is distinct from the golden angle
$\simeq137.5^{\circ}$ of phyllotaxis, which plays no role here.}  Numerically $\tilt_\phigold\simeq31.7^{\circ}$.  \rev{That value
lies $0.3^{\circ}$ from the sampled tilt $32^{\circ}$, a separation well
below the spacing of the tilt grid [Eq.~\eqref{eq:tiltgrid}] and below
any angular resolution these data support, so the two are not
distinguished in what follows: $32^{\circ}$ is taken as $\tilt_\phigold$
in the analyses below, and in the cells that sample $31.7^{\circ}$ that
angle is used in its place.  Either way the variance at the golden tilt
is read off the grid rather than interpolated.}  Its
performance is quantified by the relative variance penalty
$v(\tilt_\phigold)/v_{\min}$, the variance at the golden tilt
normalized by the smallest variance sampled in the same cell.

  \vspace{-0.5cm}

\section{Simulation methods}
\label{sec:model}

We sample the system by Monte Carlo at fixed particle number and sphere
radius, with each rod orientation held rigidly
fixed to the director field of Eq.~\eqref{eq:tiltfield}.  The rods are
hard, so the acceptance rule is purely geometric---a move is rejected
on overlap and accepted otherwise.  The
protocol (hard-core exclusion, displacement moves that translate rods
without rotating them, dense preparation, and production sampling) is
that of the
companion Communication and the earlier locked-rod
studies~\cite{AllahyarovSmC,Allahyarov2017,Allahyarov2018}.

\new{We restate the protocol here in the detail needed to reproduce the
present measurements.  One Monte Carlo sweep consists of $N$ single-rod
microsteps.  Each microstep proposes moving one rod center to a new
anchor drawn uniformly in a spherical cap of half-opening $\psi$ about
its current position, after which the orientation is snapped exactly to
the director $\nhat(\theta,\phi)$ of Eq.~\eqref{eq:tiltfield} at the new
anchor; rotational moves are absent, so the orientational energy is
constant and acceptance reduces to the hard-core gate.  Overlap is tested
with the Vega--Lago segment-to-segment minimum
distance~\cite{VegaLago1994} against a Verlet neighbor list built with
cutoff $\Lrod+1.5\Drod$.  The cap opening $\psi$ is auto-tuned during
equilibration toward an acceptance rate near $30\%$
($\psi\!\to\!0.85\psi$ below $20\%$, $\psi\!\to\!1.15\psi$ above $45\%$,
clamped to $[0.005,1.0]$~rad).  Dense configurations are prepared by
Fibonacci-spiral seeding on an enlarged sphere with thinner rods followed
by a sphere-shrink and diameter-regrowth ramp.  Each panel is then
equilibrated for \upd{$3000$} sweeps and sampled over $40\,000$ production
sweeps at intervals of ten sweeps, so that the polar marginal reported
below is an average over $4000$ recorded profiles.
The profiles are
accumulated in equal-width colatitude bins  $2.5^{\circ}$ wide.
\rev{The variance and the equator excess are computed from the average
over all recorded profiles, the first-crossing angle from the average over
the second half of them.}}

The study spans fifteen $(\Rsph/\Drod,\Lrod/\Drod)$ geometries
($\Rsph\in\{10,20,30,40\}\Drod$, $\Lrod\in\{3,5,8,10\}\Drod$,
excluding the geometrically frustrated corner $(10,10)\Drod$) at
nominal projected-area fractions $\eta\in\{0.2,0.4,0.6,0.75\}$, where
\begin{equation}
  \eta=\frac{N A_{\rm rod}}{4\pi\Rsph^2},
  \qquad
  A_{\rm rod}=\Lrod\Drod+\frac{\pi\Drod^2}{4}
  \label{eq:eta}
\end{equation}
with $N$ the number of rods and $A_{\rm rod}$ the projected area of a
single spherocylinder (cylinder plus two hemispherical caps).
\rev{Equation~\eqref{eq:eta} is a nominal projected-area fraction.  It
counts the projected areas of $N$ rods against the area of the host sphere
and takes no account of the standoff of a straight rod from a curved
surface discussed in Sec.~\ref{sec:scaling}, so $\eta$ is a control
variable for the number of rods per unit host area and not a measured
surface coverage; the distinction grows with $\Lrod/\Rsph$.
$N$ runs from $29$ in the smallest
cell to $3984$ in the largest.
}

\new{Three terms are used for the three groupings of the data set.  A
cohort is one simulated $(\Rsph,\Lrod)$ pair, a cell is one
geometry--density combination $(\Rsph,\Lrod,\eta)$, and a panel is one
geometry--density--tilt combination.}
Each cell is sampled at the locked tilts
\begin{equation}
  \tilt\in
  \{0,22.5,28,31.7,32,35,38,42,45,55,67.5,90\}^{\circ}
  \label{eq:tiltgrid}
\end{equation}
The grid is deliberately denser between
$28^{\circ}$ and $45^{\circ}$, where the sampled minima fall, than at the
two extremes. $v_{\min}$ in a cell is the minimum over the angles that cell
actually samples, so cells with unequal grids are not strictly comparable at
that level of detail\har{.  Where a quantity is averaged over cohorts at
fixed tilt in Sec.~\ref{sec:core}, only the angles common to all cohorts
enter that average}.

\new{The $\eta=0.75$ geometries are shared with the companion
Communication~\cite{AllahyarovSmC} and were re-run for the present work
with a longer production stage.  They enter here only through the
polar-marginal observables of Sec.~\ref{sec:theory}.}

\begin{figure}[!t]
\centering
\includegraphics[width=0.95\linewidth]{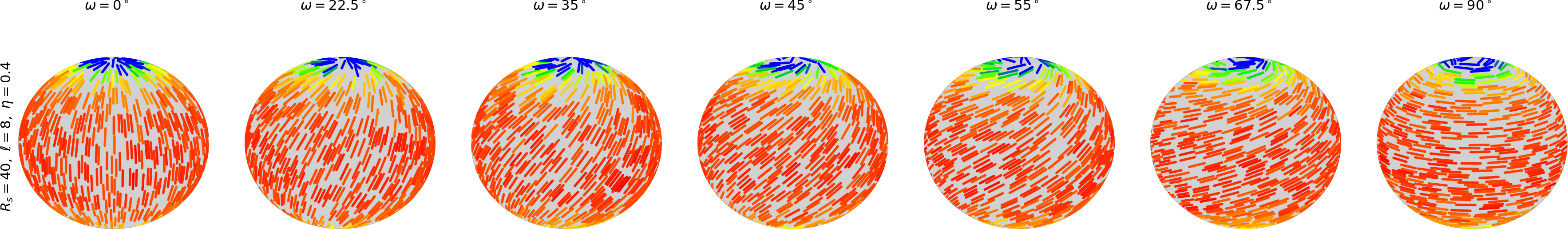}\\[1pt]
\includegraphics[width=0.95\linewidth]{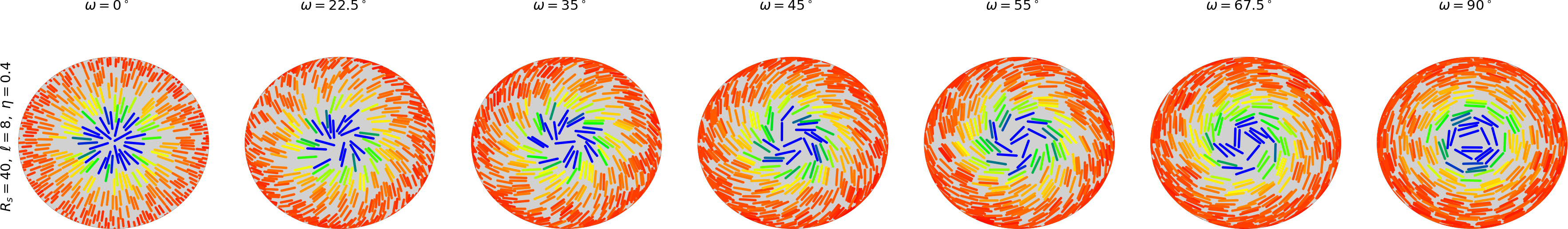}\\[1pt]
\includegraphics[width=0.95\linewidth]{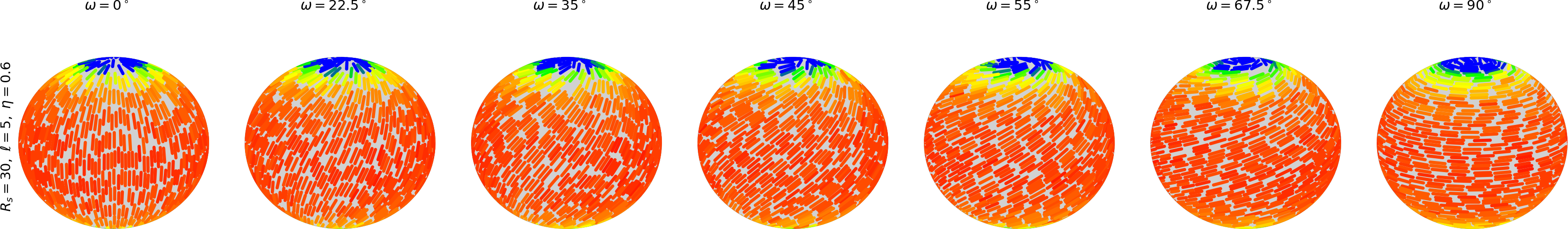}\\[1pt]
\includegraphics[width=0.95\linewidth]{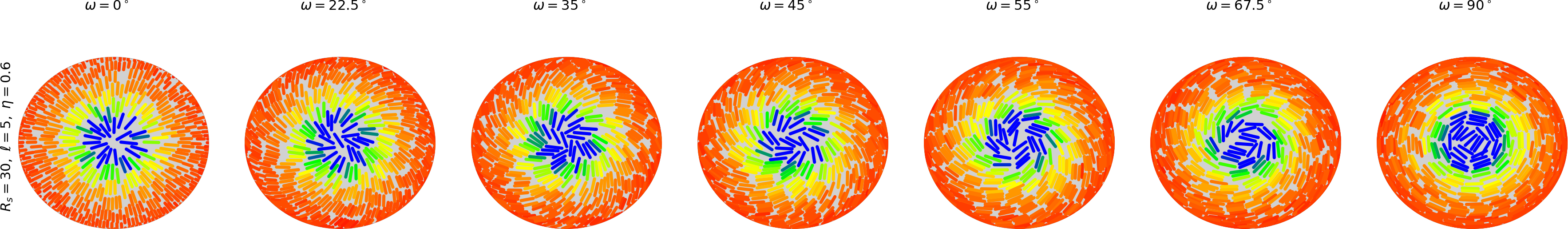}
\caption{Simulation snapshots at the locked tilts $\tilt=0$,
$22.5^{\circ}$, $35^{\circ}$, $45^{\circ}$, $55^{\circ}$,
$67.5^{\circ}$, $90^{\circ}$ (left to right) for two representative
cells at intermediate densities.  \emph{Top two rows:}
$(\Rsph,\Lrod)=(40,8)\Drod$ at $\eta=0.4$, side view (first row) and
view along the polar axis (second row).  \emph{Bottom two rows:} the
same views for $(30,5)\Drod$ at $\eta=0.6$.  Rods are colored by their
local nematic order parameter \upd{$S_i$, defined in the text}.
\rev{The color scale covers the full range $-1/2\le S_i\le1$ but is not
linear in $S_i$: it saturates at blue for $S_i\le0.7$ and runs from blue
through green, yellow and orange to red between $S_i=0.7$ and $S_i=1$, so
differences among poorly aligned rods are not resolved by color.}
\har{The imposed field of Eq.~\eqref{eq:tiltfield} does carry an exact
up-down symmetry, but it is the two-fold rotation about an equatorial axis,
$(\theta,\phi)\to(\pi-\theta,-\phi)$, and not a reflection in the equatorial
plane: at intermediate tilts the loxodromic texture is chiral, and
reflection maps $\omega$ to $-\omega$, the field of opposite handedness.
The ensemble inherits that rotational symmetry, so the two poles are
statistically equivalent, but each panel is a single instantaneous
configuration rather than an average, so its two hemispheres need not look
alike.  The symmetry is recovered in the averaged quantities: the profiles
are accumulated over thousands of configurations, and $\theta_{1/2}$ is
obtained by averaging the crossings found at the two poles.}}
\label{fig:snapshots}
\end{figure}

Figure~\ref{fig:snapshots} shows representative snapshots at two
intermediate densities.
\upd{Rods are colored by their local nematic order parameter, the average
of the second Legendre polynomial $P_2(z)=(3z^2-1)/2$ over the neighbors
of a rod,
$S_i=\langle P_2(\hat{\mathbf u}_i\!\cdot\!\hat{\mathbf u}_j)
\rangle_{j\in\mathcal N_i}$, where $\hat{\mathbf u}_i$ is the axis of rod
$i$ and $\mathcal N_i$ collects the rods whose centers lie within a fixed
cutoff of the center of rod $i$ ($10\Drod$ for the $(40,8)\Drod$ rows and
$7.5\Drod$ for the $(30,5)\Drod$ rows).  A value $S_i=1$ means the rod is
parallel to all its neighbors, $S_i=0$ means they are randomly oriented,
and $S_i=-1/2$ means they are perpendicular to it.}
\upd{The locked tilt reorganizes the coverage
visibly.  In the polar views of the second and fourth rows the director
turns from radial spokes that converge on the singularity under meridian
locking into closed rings that encircle it under latitude locking, and the
poorly ordered neighborhood of each singularity tightens between meridian
locking and the intermediate tilts.}

\FloatBarrier

\section{Results}
\label{sec:equator}

\new{\subsection{Equator-heavy coverage and the state of the
singularities}
\label{sec:profiles}}

\begin{figure}[!t]
  \centering
  \includegraphics[width=\linewidth]{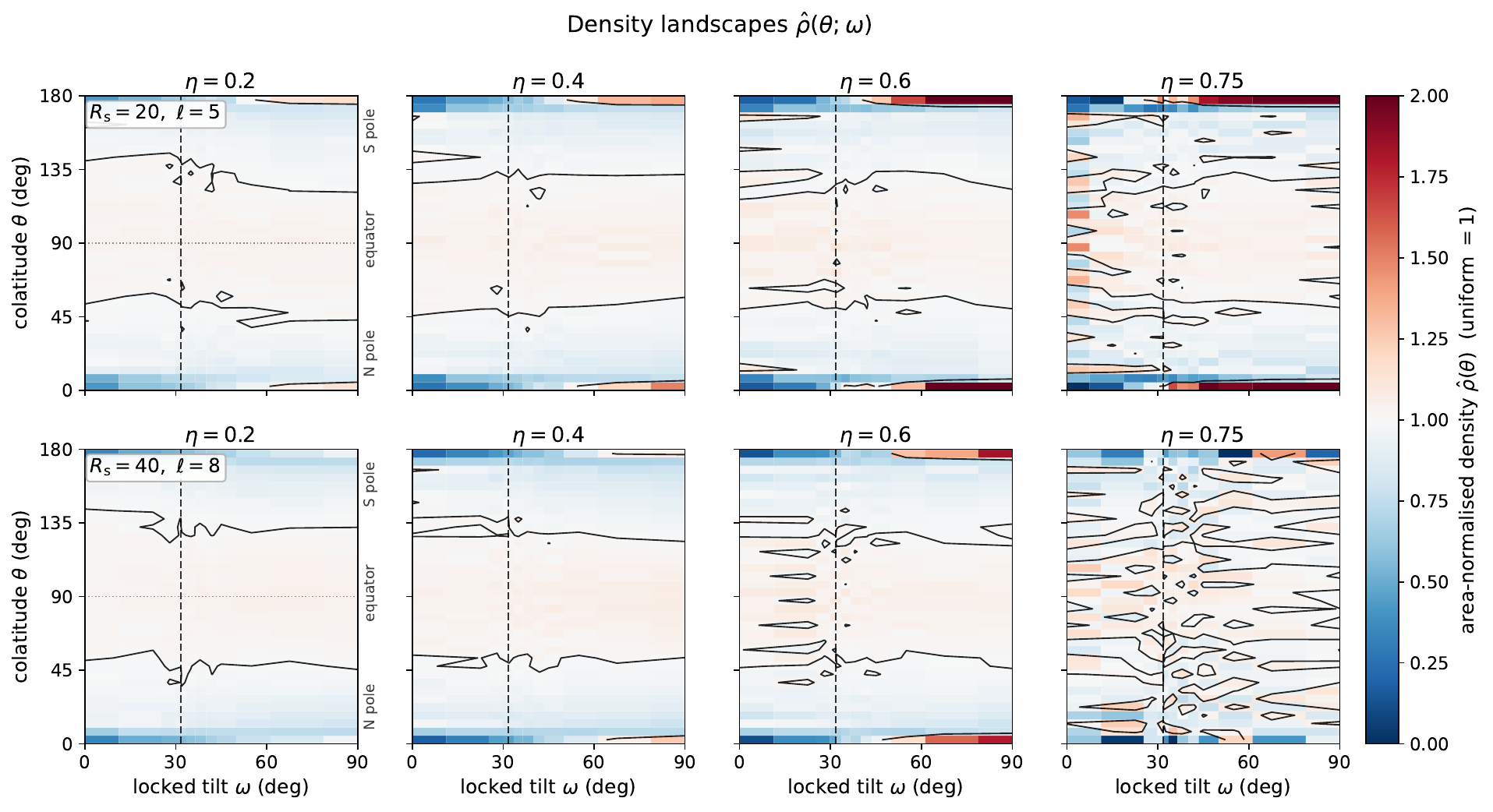}
  \caption{Area-normalized polar density $\hat\rho(\theta;\tilt)$ for
  $(\Rsph,\Lrod)=(20,5)\Drod$ (top) and $(40,8)\Drod$ (bottom) at the four
  nominal projected-area fractions.  Blue and red mark $\hat\rho<1$ and
  $\hat\rho>1$, and the black contour is $\hat\rho=1$.  \new{The color
  scale saturates at $\hat\rho=2$, so the narrow pole-adjacent spikes
  discussed in the text, which reach $\hat\rho\simeq6.5$, all appear as
  the darkest red and their heights cannot be read off the panels.}
  The dashed line
  marks the sampled tilt-grid angle $31.7^{\circ}$.
  Colatitude $\theta=0$ and $180^{\circ}$ are the poles, and
  $\theta=90^{\circ}$ is the equator.}
  \label{fig:atlas}
\end{figure}

\har{Since $\Delta_{\rm eq}$ is one of the three observables, we say how it
is distributed before turning to individual profiles.  The equator excess
is positive in all but a handful of panels, so the coverage is denser
around the equator than over the polar caps, and its typical value is
$\Delta_{\rm eq}\simeq0.15$ on the scale where uniform coverage is unity: a
modest equator-heaviness, consistent with the regional ratio reported in
Sec.~\ref{sec:jcost}.  Every panel at the two lower coverages is
equator-heavy.  What is notable is how little the magnitude responds to
density, staying close to that typical value as the sphere fills and
dipping only slightly at the highest coverage, so the excess neither grows
nor disappears with filling.  It responds instead to the tilt, falling from
meridian to latitude locking, the same ordering the variance gives in
Sec.~\ref{sec:tilt}.  The few negative panels are coverage
inversions, confined to dense and strongly tilted cases, and are discussed
below.}
Figure~\ref{fig:atlas} shows the polar marginal $\hat\rho(\theta;\tilt)$
for two representative geometries, $(20,5)\Drod$ and $(40,8)\Drod$, at
all four densities.  The dominant feature is the state of the two
singular points, and it is systematic across the whole figure rather
than particular to any one density.  Under meridian locking the bin
touching each pole is strongly depleted at every density in both
geometries, and the depletion deepens as the sphere is filled: the
pole-bin density falls from $\hat\rho\simeq0.42$ at $\eta=0.2$ to
\new{$0.15$} at $\eta=0.6$ for $(20,5)\Drod$, and from \new{$0.30$ to
$0.07$} over
the same range for $(40,8)\Drod$.  Locking the director along meridians
therefore empties the neighborhood of the singularities, which is the
blue wedge at $\theta=0^{\circ}$ and $180^{\circ}$ on the left edge of
every panel.

Turning the director from meridian toward latitude fills the poles, and
it does so at every density.  The tilt at which the pole ceases to be
depleted, meaning the smallest sampled $\tilt$ at which the pole-bin
density reaches the uniform value, moves steadily to lower angles as
the density rises.  For $(20,5)\Drod$ it lies at $\tilt=67.5^{\circ}$,
$55^{\circ}$, \new{$35^{\circ}$ and $31.7^{\circ}$} at $\eta=0.2$, $0.4$, $0.6$
and $0.75$ respectively, and for $(40,8)\Drod$ \new{it moves from
$90^{\circ}$ at $\eta=0.2$ to $67.5^{\circ}$ and $45^{\circ}$ at
$\eta=0.4$ and $0.6$}.  Denser packings thus need
less reorientation of the director before rods begin to occupy the
singularities.

The pole-touching accumulation is not the same thing as a coverage
inversion, and the distinction is one of scale.  The bin adjacent to a
pole carries under $3\%$ of the solid angle of the polar cap
$\theta<30^{\circ}$ entering Eq.~\eqref{eq:deq}, and the annulus
surrounding it stays depleted, so the area-weighted cap average remains
below unity in every panel of Figure~\ref{fig:atlas}, in the range
\new{$0.84$}--$0.98$, while the equatorial belt stays close to unity, between
$0.99$ and $1.06$.  Consequently $\Delta_{\rm eq}>0$ throughout both
geometries, with a smallest value of \upd{$+0.010$}.  A sign change of $\Delta_{\rm eq}$
requires the polar enrichment to be broad rather than confined to the
singularity: in the geometries that do invert the enrichment extends
over the inner $15^{\circ}$ of the cap, with $\hat\rho\simeq6.5$,
\new{$4.4$ and $2.4$} in successive bins away from the pole, so that it survives
the area weighting.  All inversions occur at $\eta\ge0.6$ and
$\tilt\ge67.5^{\circ}$, and the most negative value recorded is
$\Delta_{\rm eq}=-0.110$.  At the highest density the monotonic filling
of the poles with tilt is lost for $(40,8)\Drod$, where the pole-bin
density \new{fluctuates between zero and $1.2$} \upd{because of the
smectic layering that forms at this coverage~\cite{AllahyarovSmC}}.

\begin{figure}[!t]
  \centering
  \includegraphics[width=0.82\linewidth]{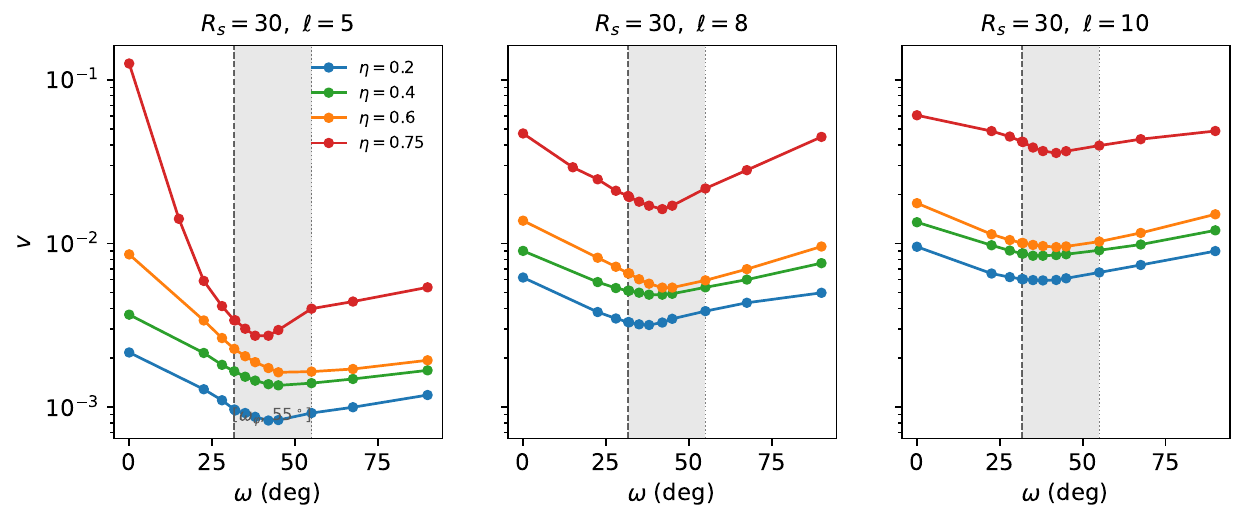}
  \caption{Polar-marginal variance $v$ versus locked tilt for
  $\Rsph=30\Drod$ and $\Lrod=5,8,10\Drod$.  \new{The shaded interval runs
  from the golden tilt $\tilt_\phigold\simeq31.7^{\circ}$ of
  Eq.~\eqref{eq:golden} to $55^{\circ}$.  It is a reference plotting band
  that brackets where the sampled minima fall \aud{in the great majority of
  cells}.}}
  \label{fig:vlandscapes}
\end{figure}

\new{\subsection{Tilt dependence of the coverage uniformity}
\label{sec:tilt}}

Figure~\ref{fig:vlandscapes} displays the variance curves $v(\tilt)$
for the three $\Rsph=30\Drod$ cohorts.
\new{In each of them the variance is largest at meridian locking, drops
as the director is turned, and passes through a shallow
minimum at an intermediate angle.
\upd{How far it drops depends on the cohort: the meridian variance
exceeds the smallest sampled value by little more than half for the
longest rods at the two lower coverages, and by more than an order of
magnitude in the densest cell of the shortest-rod cohort.}
The tilt that minimizes the variance
falls inside the shaded reference band from $31.7^{\circ}$ to
$55^{\circ}$ in the great majority of cells and never at meridian
locking.  The remaining cohorts behave the same way.
\rev{\aud{Because that minimum is shallow and the minimizing angle moves
from cell to cell, the physically meaningful statement is about the
interval and not about its interior:} the polar-marginal variance is small
over an
intermediate range
of locked angles, which corresponds to an approximately homogeneous
density distribution.  \aud{Inside the interval} the variance changes by
about half
in the typical cell, against a factor of two to three across the full tilt
grid and far more in the densest cells.
}}

\begin{figure}[!t]
  \centering
  \begin{minipage}[t]{0.47\linewidth}
    \centering
    \textbf{(a)}\\[1pt]
    \includegraphics[width=\linewidth]{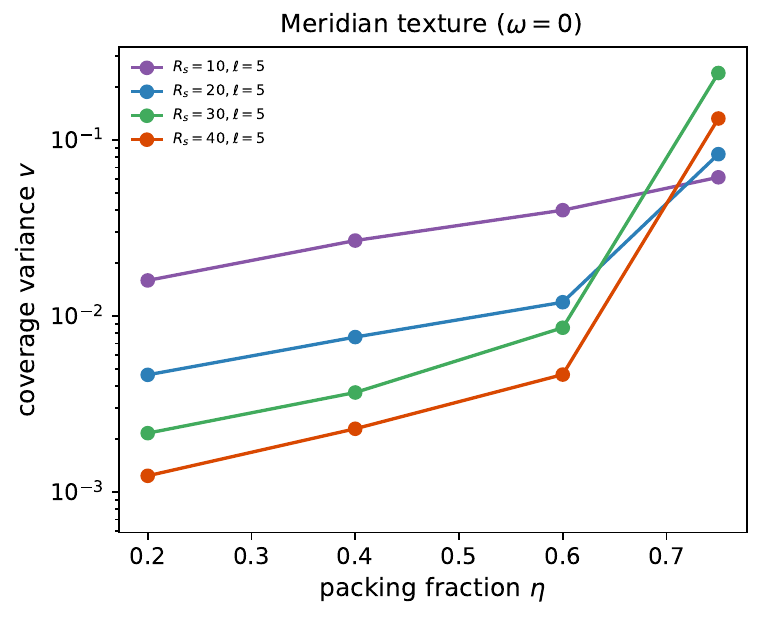}
  \end{minipage}\hfill
  \begin{minipage}[t]{0.47\linewidth}
    \centering
    \textbf{(b)}\\[1pt]
    \includegraphics[width=\linewidth]{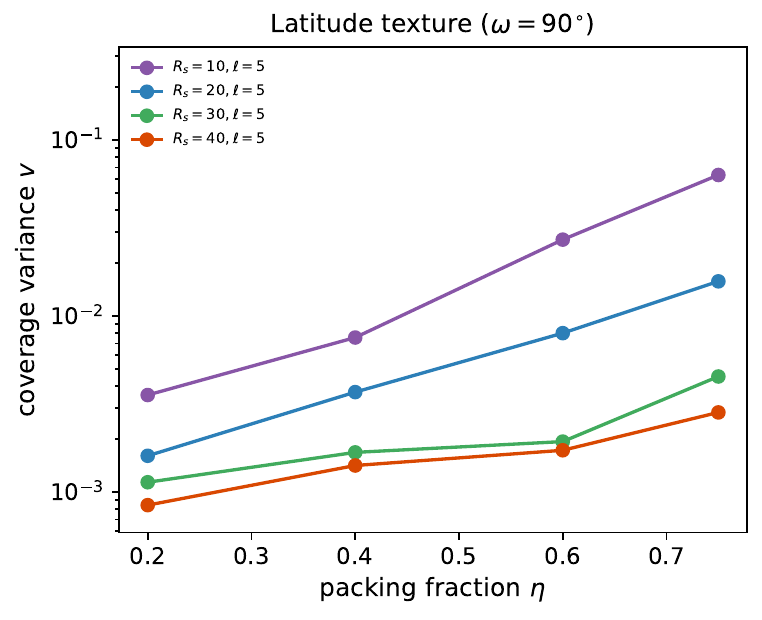}
  \end{minipage}
  \par\smallskip
  \begin{minipage}[t]{0.47\linewidth}
    \centering
    \textbf{(c)}\\[1pt]
    \includegraphics[width=\linewidth]{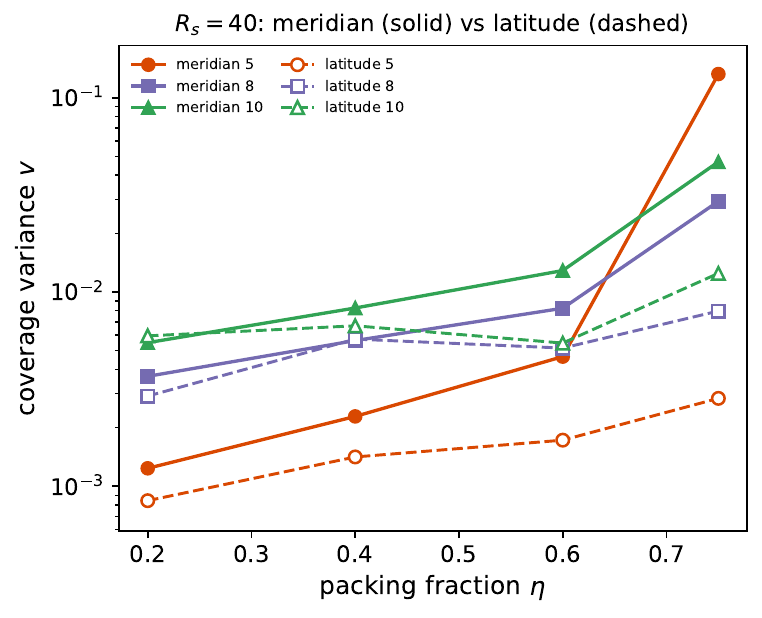}
  \end{minipage}\hfill
  \begin{minipage}[t]{0.47\linewidth}
    \centering
    \textbf{(d)}\\[1pt]
    \includegraphics[width=\linewidth]{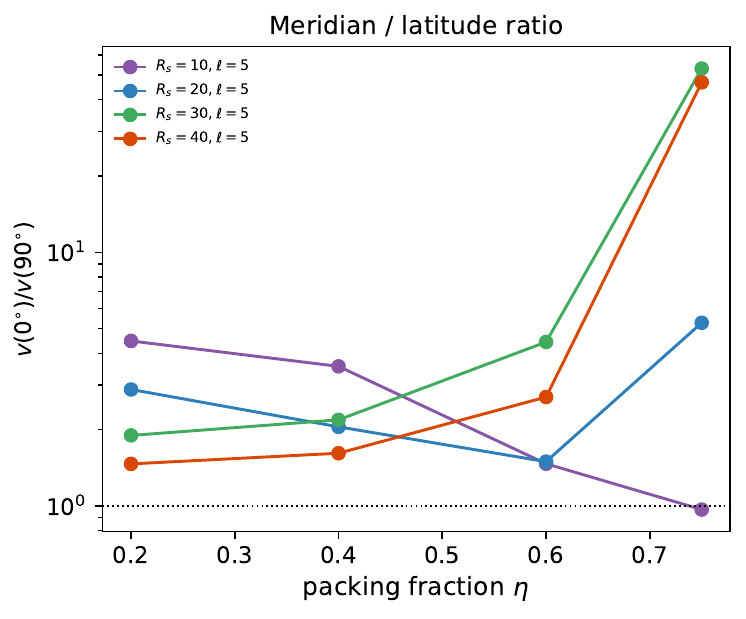}
  \end{minipage}
  \caption{Comparison of the two extreme tilts, meridian ($\tilt=0$)
  and latitude ($\tilt=90^{\circ}$), by polar-marginal variance.
  (a)~Meridian and (b)~latitude locking for
  fixed $\Lrod=5\Drod$.  (c)~The two extreme tilts at $\Rsph=40\Drod$ for
  $\Lrod=5,8,10\Drod$.  (d)~Ratio $v(0^{\circ})/v(90^{\circ})$
  \upd{for the same $\Lrod=5\Drod$ cohorts as in panels (a) and (b)},
  where the dotted line marks equal variances at the two extreme tilts.}
  \label{fig:flip}
\end{figure}

Figure~\ref{fig:flip} compares the two extreme tilts directly.
Panels (a) and (b) show the polar-marginal variance under meridian and
latitude locking, respectively, for $\Lrod=5\Drod$ at the four sphere
radii: in both textures $v$ falls as the density decreases, but the
meridian curves lie above the latitude ones in nearly every case.
Panel (c) makes the same comparison at $\Rsph=40\Drod$ for
$\Lrod=5,8,10\Drod$ (meridian solid, latitude dashed), showing that the
meridian variance exceeds the latitude variance across rod lengths.
Panel (d) quantifies this through the ratio $v(0^{\circ})/v(90^{\circ})$
of the meridian to the latitude variance; a value above unity means
that meridian locking carries the larger variance.  \new{Across the whole
data set the ratio lies above unity in nearly every cell, so meridian
locking is the less uniform of the two extreme tilts as a rule rather
than as a tendency.}


\new{\subsection{The ratio-symmetric imbalance measure}
\label{sec:jcost}}

\upd{The two quantities used in the preceding subsections keep very
different amounts of the same profile.  The variance $v$ of
Eq.~\eqref{eq:variance} weights every bin, while the equator excess
$\Delta_{\rm eq}$ of Eq.~\eqref{eq:deq} keeps only the difference between
the two regional means.  The imbalance measure $\Jcost$ of
Eq.~\eqref{eq:Jx} is built from those same two means but keeps only their
ratio $x$ of Eq.~\eqref{eq:xratio}.  We apply it to the simulations here
for the first time, and we ask three things of it.}

\upd{The first is whether the two quantities agree on which panels have
the more even coverage.  At a fixed density we sort all panels from the
smallest variance $v$ to the largest, sort the same panels a second time
from the smallest imbalance $\Jcost$ to the largest, and ask how closely
the two sequences follow one another.  The Spearman correlation between
them \upd{\cite{Spearman1904,NumericalRecipes}}, which equals one if the
two sequences are identical and zero if they
are unrelated, is $0.96$, $0.93$, $0.77$ and $0.44$ at
$\eta=0.2$, $0.4$, $0.6$ and $0.75$.  At the two lower densities the
sequences nearly coincide, so a panel with a small equator-to-pole
imbalance \aud{measured by $\Jcost$}
is also a panel with a small variance $v$.  The uniformity is then
decided by \aud{the ratio of the two regional means alone}, and the shape of the
profile inside each region matters little.  The agreement between $\Jcost$ and $v$ weakens as the
sphere fills and is poor at the highest density, where the profile changes
from one bin to the next in a way that a ratio of two regional means 
cannot see.  \rev{Both quantities are computed from the same averaged
profile, so their agreement is not an independent test of anything. What it
measures is how much of the tilt dependence of $v$ survives the reduction of
that profile to two regional means.}  We therefore use $\Jcost$ as a
diagnostic at low and intermediate coverage, and we do not treat it as a
quantity to be minimized.}

\upd{The second is how large the imbalance is.  Over the whole data set
the median $\Jcost$ is $0.012$, the central half of the panels lie
between $0.006$ and $0.022$, and the median ratio is $x=1.16$.  The
imposed tilt therefore leaves the equatorial belt about sixteen percent
denser than the polar caps, which agrees with the equator-heavy profiles
of Sec.~\ref{sec:profiles}.  Between the two extreme tilts the median
$\Jcost$ falls from $0.017$ at meridian locking to $0.008$ at latitude
locking, the same direction as the variance comparison of
Figure~\ref{fig:flip}.  The two measures do not, however, reach their
smallest values at the same tilt.  At the variance-minimizing tilt of each
cell the median $\Jcost$ is $0.011$, above its latitude value, because
balancing the two regional means is not the same as flattening the profile
within them.  We report the difference because it marks the limit of what
a measure built on two regional means can do.}

\upd{The third is where the rare coverage inversions sit.}  \rev{This is
the one question on which the two-region description adds something the
variance cannot supply, because $v$ does not distinguish the sign of the
equator--pole contrast.}  \upd{Every
panel with $\Delta_{\rm eq}<0$ has \rev{a regional ratio $x$ between
$0.90$ and $1.00$, equivalently $\Jcost\le0.006$}, an imbalance below the
median of the data set.  The
sign changes therefore occur where the equatorial and polar means are
nearly equal, and not in strongly pole-heavy coverage.
}

\new{\subsection{The golden\pit{-ratio} \pit{tilt as a} reference}
\label{sec:goldenref}}

\begin{figure}[!t]
  \centering
  \includegraphics[width=0.72\linewidth]{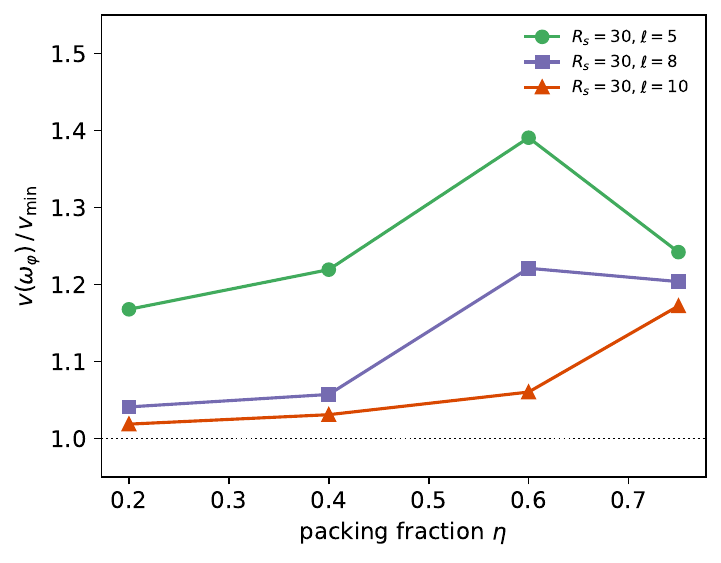}
  \caption{Relative polar-marginal variance penalty
  $v(\tilt_\phigold)/v_{\min}$ of the golden tilt for the
  $\Rsph=30\Drod$ cohorts of Figure~\ref{fig:vlandscapes}, that is the
  variance at $\tilt_\phigold\simeq31.7^{\circ}$ divided by the smallest
  variance sampled in the same cell,
  \upd{plotted against the projected-area fraction.  The dotted line
  marks unity, the value the penalty would take if the golden tilt were
  itself the best sampled choice.}
  }
  \label{fig:golden}
\end{figure}

Figure~\ref{fig:golden} shows the relative penalty
$v(\tilt_\phigold)/v_{\min}$ of Sec.~\ref{sec:theory} for the
$\Rsph=30\Drod$ cohorts, \aud{which measures what is given up by locking at
the golden tilt rather than at the tilt that minimizes the variance in that
cell.}

At every density the penalty falls as the rods lengthen.
The longest rods therefore stay \upd{at or below} $1.06$ at $\eta\le0.6$
and below $1.18$ at the highest density, whereas the shortest reach
$1.391$.
Locking the director at the golden tilt is thus nearly free for long
rods and becomes steadily more costly as the rods are shortened.



\begin{figure}[!t]
  \centering
  \begin{minipage}[t]{0.49\linewidth}
    \centering
    \textbf{(a)}\\[1pt]
    \includegraphics[width=\linewidth]{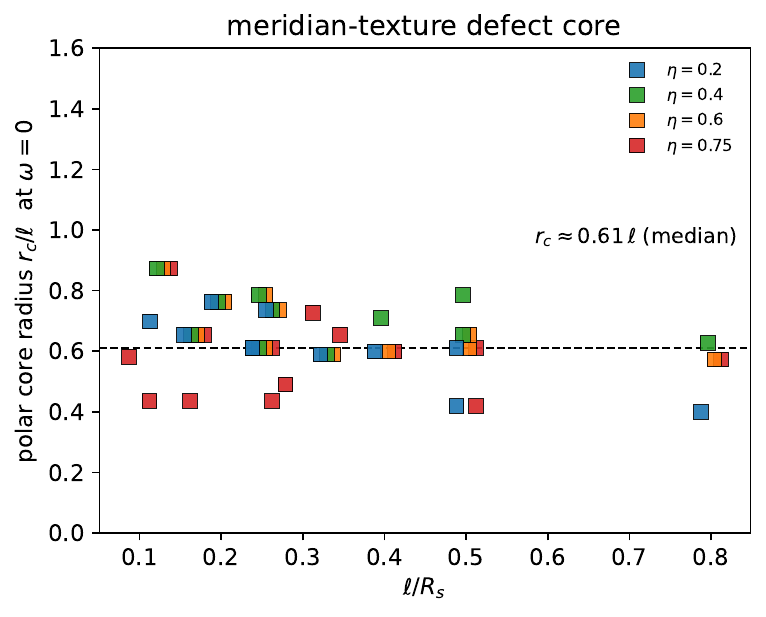}
  \end{minipage}\hfill
  \begin{minipage}[t]{0.49\linewidth}
    \centering
    \textbf{(b)}\\[1pt]
    \includegraphics[width=\linewidth]{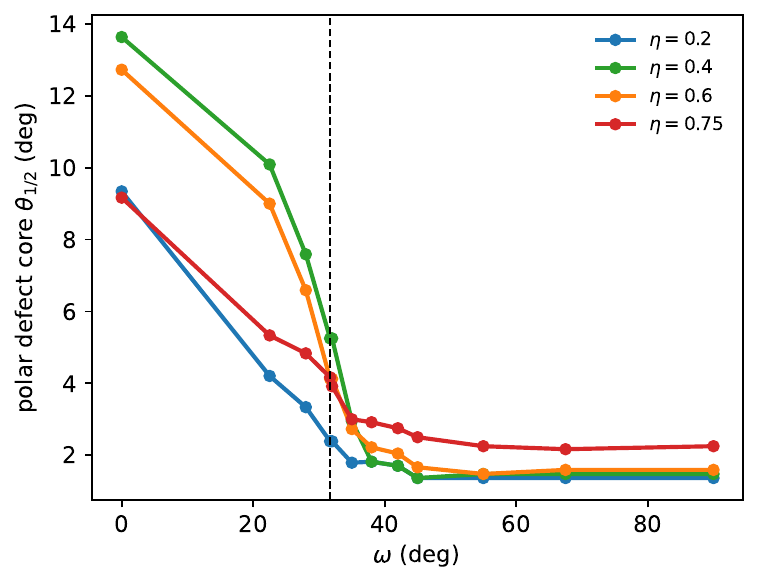}
  \end{minipage}
  \caption{\aud{Width of the depleted region around each singularity.
  (a)~Meridian first-crossing radii $r_c/\Lrod$ versus $\Lrod/\Rsph$.}
  \upd{The ordinate, labelled polar core radius in the panel, is the
  width of the depleted region as defined in Sec.~\ref{sec:core}.}
  \new{In panel \upd{(a)} a small horizontal offset is applied to the
  markers of
  each density so that scatter points which would otherwise coincide
  remain distinguishable; the offset affects the plotted abscissa only and
  not the measured value of $r_c/\Lrod$.}
  \aud{(b)~Unweighted cohort means of the first-crossing estimator
  $\theta_{1/2}$ versus locked tilt for each density, where the dashed line
  marks the $31.7^{\circ}$ reference angle.}
  \upd{The ordinate label
  polar defect core denotes this estimator, the angular radius of the
  depleted region defined in the text.}}
  \label{fig:cores}
\end{figure}

\new{\subsection{Locked tilt controls the width of the depleted region}
\label{sec:core}}

\new{We turn to the first of the two main results.  The variance $v$
measures how uneven the coverage is but says nothing about where the
unevenness sits.
The polar first-crossing angle $\theta_{1/2}$ of Sec.~\ref{sec:theory}
localizes it, since it is the angular radius of the region around a
singularity within which the coverage has fallen to half the uniform
value.  We refer to this region as the depleted region and to its arc
radius $r_c=\Rsph\theta_{1/2}$ as its width.  \aud{Panels (a) and (b) of
Figure~\ref{fig:cores}} show, respectively, how wide it is under meridian
locking and how it responds to the tilt.}

Figure~\ref{fig:cores}(a) shows the meridian first-crossing radii
$r_c/\Lrod$ versus $\Lrod/\Rsph$.  The points cluster near the dashed
line at $r_c/\Lrod=0.61$, the median over all meridian panels, with
considerable geometry- and density-dependent scatter.
\new{The content of that plot is that the width of the depleted region
under meridian locking is set \rev{principally} by the rod length: the
central half of the panels lies between $0.59\Lrod$ and $0.73\Lrod$ while
the host radius varies by a factor of four, the rod length by a factor of
three, and the coverage from $\eta=0.2$ to $0.75$.  The collapse is not
exact,
so $0.61\Lrod$ is a central value rather than a
strict law; but the rod length, not the curvature radius and not the
coverage, is what fixes the scale.  This is the natural outcome for a
depletion driven by rigid-rod exclusion, because $\Lrod$ is the only
length the rod itself brings to the singularity.}

Figure~\ref{fig:cores}(b) resolves the same estimator in tilt, showing the
unweighted cohort means of the first-crossing angle $\theta_{1/2}$
(Sec.~\ref{sec:theory}) against the locked tilt.  Every density follows
the same two-stage course.  Under meridian locking the depleted region
around each singularity is wide, $\theta_{1/2}$ lying between
\new{$9.2^{\circ}$} and $13.6^{\circ}$, and it contracts steeply as the
director is turned away from the meridian.  The contraction then stops:
beyond roughly $38^{\circ}$ the curves run flat onto a plateau of
\new{$1.4^{\circ}$ to $2.9^{\circ}$}, so that latitude locking confines the
depleted region to the innermost bins of the grid and further tilting
buys nothing.
\rev{The height of that plateau is set by the angular resolution rather
than measured: at every tilt beyond $38^{\circ}$ the crossing already falls
in the first histogram bin in most panels, so the plateau is an upper bound
on the width fixed by the bin width of Sec.~\ref{sec:model}, and the
contraction between the two extreme tilts is correspondingly a lower bound
on the true contraction.}
\new{The size of the effect is the reason we treat it as a main result.
Between the two extreme tilts the width of the depleted region falls by
\rev{at least} a factor \rev{of $4.1$ to $9.2$} depending on the coverage,
and in arc
length the median crossing radius drops from $0.61\Lrod$ at meridian
locking to $0.09\Lrod$ at latitude locking, a contraction of nearly an
order of magnitude.  \pit{The latter figure coincides with the first-bin
floor in most of the cells that enter it, so it is an upper bound fixed by
the resolution and not a measured width.}  Topology fixes that the singularities exist and
Figure~\ref{fig:cores}(a) fixes the width \mil{that packing produces around
them} when the director runs
along meridians, but \mil{that width is not itself a topological quantity}: it is set
by the locked angle, and it is adjustable over almost a decade at every
coverage studied.  A contraction of this size, reproduced at all four
coverages, is the largest and most systematic effect the locked tilt
produces anywhere in the data set.}

\pit{The two-stage course can be read against the sampled angle
$\tilt_\phigold$.}  The golden\pit{-ratio} tilt marked by the dashed line falls close to the
boundary
between these two stages.  By $\tilt_\phigold\simeq31.7^{\circ}$ the
cohort means have already completed most of their total contraction,
\upd{$87\%$ at $\eta=0.2$ and $69$ to $77\%$ at the three higher
densities,}
and the residual descent is spent within the following few degrees.
Locking anywhere beyond the golden\pit{-ratio} tilt therefore leaves \upd{that width}
\pit{unchanged as far as the estimator resolves}, while locking below it forfeits a large part of
the available contraction.
\new{The coincidence is approximate rather than sharp: the slope collapses
between $35^{\circ}$ and $38^{\circ}$, a few degrees above
$\tilt_\phigold$, and at $\eta=0.4$ roughly one third of the contraction
still remains at the golden tilt.}
\pit{Two further qualifications belong with the percentages above.  They
are fractions of the contraction the binned estimator can follow, not of
the true contraction, because the plateau against which they are measured
is itself an upper bound.  And $\tilt_\phigold$ is one of the angles the
grid happens to sample near the crossover; the proximity is an observation
about the grid, not a threshold the model predicts.}


\begin{figure}[!t]
  \centering
  \includegraphics[width=0.72\linewidth]{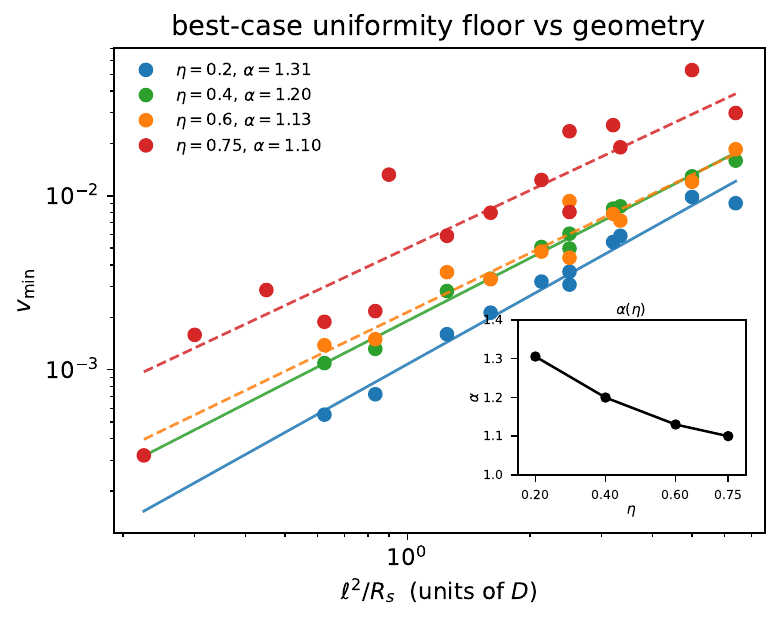}
  \caption{\aud{Lowest polar-marginal variance on each available tilt
  grid, $v_{\min}$, versus $\Lrod^2/(\Rsph\Drod)$ (one point per cell), on
  logarithmic axes.}  \pit{The abscissa, written $\ell^{2}/R_s$ in units of
  $D$ in the panel, is that dimensionless group.}  \aud{The lines are the least-squares regressions of
  $\log v_{\min}$ on $\log[\Lrod^{2}/(\Rsph\Drod)]$ at fixed density
  described in Sec.~\ref{sec:scaling}, the fitted exponent $\alpha$ of each
  is given in the legend, and the inset shows those exponents against
  density.}
  }
  \label{fig:floor}
\end{figure}

\new{\subsection{Geometric scaling of the \rev{best sampled} uniformity}
\label{sec:scaling}}

\new{We turn to the second main result.}  For each cell we select the
smallest variance on its discrete tilt
grid.  Figure~\ref{fig:floor} displays the resulting minima, one
per cell, against $\Lrod^2/(\Rsph\Drod)$.  Their density-wise scaling is described
by
\begin{equation}
  v_{\min}
  \propto
  \left[\frac{\Lrod^2}{\Rsph\Drod}\right]^{\alpha(\eta)}
  \label{eq:floor}
\end{equation}

\new{Three points have to be made about Eq.~\eqref{eq:floor}: why this
combination of the three lengths, \aud{what the fit is}, and what it does and
does not assert.}

\new{The combination \mil{is a natural measure of} the geometric frustration
a rigid rod experiences on a curved host\mil{, and we claim naturalness for
it rather than uniqueness}.
\upd{The center of a rod lies on the sphere of radius $\Rsph$ and its axis
is tangent to the surface there (Sec.~\ref{sec:theory}), so the rod meets
the host at its midpoint alone and its ends are lifted clear of the
surface.  For rods short compared with the host radius that lift is
$\Lrod^{2}/(8\Rsph)$, which ranges from three hundredths of a rod diameter
for the shortest rods on the largest sphere to eight tenths of one for the
longest rods on the smallest.}  The natural
yardstick for that standoff is the
rod diameter, since a mismatch much smaller than $\Drod$ is absorbed
within the thickness of the monolayer while a mismatch of order $\Drod$
cannot be.  The dimensionless group $\Lrod^{2}/(\Rsph\Drod)$ is therefore
eight times \upd{that lift} expressed in rod diameters, and
across the fifteen geometries studied here it runs from $0.23$ to $6.4$.
Equivalently the same group factorizes as
$(\Lrod/\Drod)(\Lrod/\Rsph)$, the product of the rod aspect ratio and the
curvature ratio, so it increases both with slenderness and with
curvature.}

\mil{The standoff has a coordinate-free form that explains why the group
carries no tilt.  For a short rod laid tangent to a smooth surface along a
unit direction $\nhat$, the lift of its ends above the surface is
$h\simeq\tfrac{1}{8}\Lrod^{2}|\mathrm{II}(\nhat,\nhat)|
 =\tfrac{1}{8}\Lrod^{2}|k_{\nhat}|$, set by the second fundamental form
evaluated along the rod, that is by the normal curvature $k_{\nhat}$ in the
direction the rod points.  The sphere is umbilic, so $k_{\nhat}=1/\Rsph$ in
every tangent direction, and the standoff is the same whichever way the rod
is turned.  The extrinsic mismatch $\Lrod^{2}/(\Rsph\Drod)$ is therefore
tilt-independent as a matter of geometry, which is the sense in which
Eq.~\eqref{eq:floor} is a statement about the host and not about the
anchoring angle.  We quote the short-rod expansion rather than the exact
lift $\sqrt{\Rsph^{2}+(\Lrod/2)^{2}}-\Rsph$, which it closely tracks over
the geometries sampled, because it is the form in which the three lengths
separate.}

\new{The fit itself is an ordinary least-squares regression of
$\log v_{\min}$ on $\log[\Lrod^{2}/(\Rsph\Drod)]$ at fixed density, one
  point per cell.
  The lines
  drawn in Figure~\ref{fig:floor} are those regressions.
  All four
  fitted exponents lie between 1.1 and 1.3.
  \pit{The highest coverage is a qualified member of that set.  The
  $\eta=0.75$ cells carry the smectic layering of the companion
  Communication~\cite{AllahyarovSmC}, which structures the polar marginal
  in a way the standoff picture does not describe, and
  Sec.~\ref{sec:profiles} already reports the loss of monotonic pole
  filling there.  The scaling is therefore best read from the three lower
  coverages, with $\eta=0.75$ an extension rather than a fourth
  independent confirmation.}
}

What Eq.~\eqref{eq:floor} asserts is limited in the following  way.
  The parameter $v_{\min}$ is the smallest
variance found on a discrete grid of locked angles and not the minimum over
continuous $\tilt$, so it is an upper bound on the latter.
\har{The scaling is also an interpolation across the window of
$\Lrod^{2}/(\Rsph\Drod)$ actually sampled and must not be extrapolated to
the ends of that variable.  The needle limit makes this concrete.  Holding
$\Lrod$ and $\Rsph$ fixed and letting $\Drod\to0$ sends the group to
infinity, so Eq.~\eqref{eq:floor} taken literally would send $v_{\min}$ to
infinity as well, whereas infinitely thin rods have vanishing excluded
volume, occupy no area, and can only cover the sphere more evenly, not less.
The resolution is that $\Drod$ enters Eq.~\eqref{eq:floor} through the
standoff measured in rod diameters, a ratio that ceases to be the relevant
comparison once the rod is thinner than any other length in the problem: at
fixed particle number the coverage of Eq.~\eqref{eq:eta} then tends to zero
and the packing that produces the depletion disappears with it.  The same
caution applies at the opposite end, where the group is large because the
rods are long enough for the tangency constraint to dominate.  We therefore
read Eq.~\eqref{eq:floor} as an empirical description within the sampled
window and not as a limiting law.}
\section{Discussion and conclusions}
\label{sec:discussion}
\label{sec:conclusions}

\new{Two statements survive the survey, and they answer the two questions
raised in Sec.~\ref{sec:intro}.  The first is that the locked tilt is a
usable control on the geometry of the defect.  Topology dictates that a
tangential director field on a sphere be singular, and the coverage
responds with a depleted region around each singularity whose width under
meridian locking is set \pit{principally} by the rod length.
\aud{The tilt then determines how wide that region is.}  The practical reading is that the bare
patch a coating must tolerate is not a fixed cost of the topology but a
quantity the anchoring angle sets, with a floor that is reached well
before the latitude texture.}

\new{The second statement concerns what tilt cannot do.  Choosing the tilt
optimally still leaves a residual nonuniformity, and that residual is
governed by the geometry through $\Lrod^{2}/(\Rsph\Drod)$, the
out-of-surface mismatch of a straight rod measured in rod diameters.
  \rev{Long rods on small hosts
are therefore not made uniform in the polar marginal by any of the locked
angles sampled}, and the way to
improve the uniformity is to change the mismatch rather than the anchoring
direction.}
\rev{We emphasize that Eq.~\eqref{eq:floor} is an empirical scaling of the
  smallest sampled variance over the sampled grids and not a bound.
}

\new{Between these two, the tilt dependence of the uniformity is real but
coarse.  Meridian locking is the least uniform choice throughout, and the
variance-minimizing tilt lies at intermediate angles in the great majority
of cells.  We deliberately state that as a
band from $31.7^{\circ}$ to $55^{\circ}$ and not as an optimal angle.}
\aud{The minimum is shallow and the tilt grid is coarse, so what the data
fix is the interval and not a preferred angle inside it.}

A plausible reading of the prevalent equator-heavy marginal is
excluded-volume packing near the imposed polar
singularities~\cite{Shin2008,Smallenburg2016,Allahyarov2017}.  The
present design, however, does not isolate this mechanism from
finite-size, binning, preparation, or singularity-location effects.
\new{The ratio-symmetric measure $\Jcost$ of Eq.~\eqref{eq:Jx} adds one
constraint on that reading.  It ranks the locked angles almost exactly as
the variance does at the two lower coverages and progressively less well
as the sphere fills, so at low coverage the entire tilt dependence is
carried by the equator-to-pole ratio and not by structure within either
region.  The rare sign reversals of the equator excess all occur
\rev{where the two regional means are nearly equal}\aud{, as
Sec.~\ref{sec:jcost} shows}, so they are marginal
crossings of balance rather than a distinct pole-heavy state.}

\mil{Equations~\eqref{eq:splaybend} and~\eqref{eq:distortion} narrow the
candidate mechanism for the tilt control of the width.  Because the total
distortion of the locked field is tilt-independent, the contraction of
Figure~\ref{fig:cores}(b) cannot be read as the field becoming less
distorted; the tilt only trades splay for bend.  Of the two,
splay is the component a rod of finite length resolves, since it asks
neighboring rods to separate along their own axes, where the length that
must be accommodated is $\Lrod$, whereas bend separates them across their
axes, where it is only $\Drod$.  The reading most consistent with our data
is therefore that packing near the singularity responds to the splay rather
than to the total distortion.  The comparison applies to the first stage of
Figure~\ref{fig:cores}(b) alone.  Splay falls off as $\cos\tilt$, gradually
at first, so it can account for a contraction that begins at meridian
locking and continues through the intermediate angles, but it predicts no
plateau; the plateau in the data is the floor of the binned estimator
identified in Sec.~\ref{sec:core}, beyond which the width is not resolved
and the two cannot be compared at all.
We stress that the present design cannot test
even the restricted statement.  Each cell imposes a single field and
measures the density, and Eq.~\eqref{eq:distortion} ties splay and bend
together, so no independent variation of the two at fixed total is
available here.  Changing the host would not by itself lift the
degeneracy: at a constant locking angle the same cancellation holds on any
surface of revolution, with $\cot\theta/\Rsph$ replaced by $f'/f$ for a
profile radius $f$ differentiated with respect to meridian arc length.
Separating the two would require a tilt that varies with colatitude,
$\tilt(\theta)$, which the present protocol does not implement.  The
statement is a hypothesis the geometry suggests, not a result the
simulations establish.}

\pit{The golden-ratio tilt is an incidental marker and we claim nothing
more for it.  By the variance it is unremarkable: the penalty
$v(\tilt_\phigold)/v_{\min}$ of Figure~\ref{fig:golden} never reaches unity,
it is slight for the longest rods and largest for the shortest, and though
the angle is the sampled minimizer in a minority of cells elsewhere in the
data set, it is not so systematically.  By the width of the depleted region
it sits near the crossover between the two stages of
Figure~\ref{fig:cores}(b), so most of the contraction the estimator can
follow is already spent there.  That proximity is an observation about
where a coarse grid was sampled, on an estimator whose plateau is set by
the bin width, and not a threshold: nothing in the model predicts that the
golden ratio should select a saturation angle, and the slope collapse in
fact occurs a few degrees above $\tilt_\phigold$.  We therefore draw no
design rule from the angle.}

\rev{Three features of the model bound these conclusions, and the first is
the locking itself.  The orientations here are imposed and athermal: each
axis is held on the tangential field of Eq.~\eqref{eq:tiltfield} whatever
its neighbors do, which is the limit of infinitely strong anchoring to an
external field and removes both the orientational entropy of the rods and
the elastic relaxation a Frank free energy would perform near a
singularity.  What the results describe is therefore the packing of hard
rods around a prescribed singularity and not the self-consistent core
structure of a nematic defect\pit{, and both main results are properties of
that ensemble}.  \pit{Neither should be expected to survive partial
unlocking in the same way.}  The geometric floor is a
statement about excluded volume: it follows from the standoff of a straight
rod on a curved host, which is unchanged when the axis is free to turn
within the tangent plane, so we expect a floor of this kind to persist as
long as the rods remain tangentially anchored.  The tilt control of the
depleted width, in contrast, uses the locked angle as a handle\pit{, and} a director
free to relax would select its own angle near each singularity.
\mil{What would set that angle is less obvious than it appears.  Because
the elastic energy of the imposed field is the same at every tilt in the
one-constant approximation (Sec.~\ref{sec:theory}), elasticity by itself
gives a relaxing director no reason to prefer one angle over another.  The
selected angle would be fixed instead by the competition between the
anisotropy of the splay and bend elastic constants and the packing
contribution studied here, and the present ensemble contains neither.}
\pit{Which
of the two survives at finite anchoring the present ensemble cannot say,}
and for the same reason the design language used above is a
statement about an idealized model: a real coating carries a fluctuating
director, finite anchoring, a finite shell thickness, particle--substrate
interactions and disorder in the imposed field, none of which enter here.}

\rev{The second is resolution.  The widths of the depleted region rest on a
half-density crossing read off a binned profile, so the plateau of
Figure~\ref{fig:cores}(b) is an upper bound fixed by the bin width and the
contraction factors are lower bounds.  The third is the sampling of the
control parameters.  The tilt grid is discrete, unequal between a few cells,
and deliberately denser between $28^{\circ}$ and $45^{\circ}$ than
elsewhere, so $v_{\min}$ is the smallest sampled variance and the
low-variance band is defined by the angles actually run.  The variance
itself probes the polar marginal only and is blind to structure around a
circle of constant latitude\pit{, and that marginal counts rod centers, so
it stands as a proxy for the covered area rather than a measurement of it}.
The highest-coverage panels of
Figure~\ref{fig:atlas} lose the otherwise monotonic filling of the poles with
tilt once the smectic layering of the companion
Communication~\cite{AllahyarovSmC} sets in, so the systematic behavior is
best read off the three lower coverages.}

\rev{\har{Against that background the contribution stated in
Sec.~\ref{sec:intro} can be made specific.  What is new is the continuous
map of the polar marginal in the locked angle, and the separation it makes
possible between a defect width that the anchoring angle controls and a
uniformity floor that the geometry fixes.  Concretely, the three lower
coverages, the tilt-resolved widths of Figure~\ref{fig:cores}(b) and the
scaling of Eq.~\eqref{eq:floor} are new here, while the $\eta=0.75$
geometries are those shared with the companion
Communication~\cite{AllahyarovSmC} and are re-analyzed through the
polar-marginal observables of Sec.~\ref{sec:theory}.}}

\mil{The separation between the two results is sharpest as a prediction for
hosts that are not spheres.  The floor is tilt-free here because the sphere
is umbilic, so the normal curvature along a rod is $1/\Rsph$ whichever way
the rod points.  On a non-umbilic surface of revolution, an ellipsoid or a
torus, the meridians and parallels are the principal directions, and Euler's
formula gives the normal curvature along a rod locked at angle $\tilt$ as
$k_{\nhat}=k_{1}\cos^{2}\tilt+k_{2}\sin^{2}\tilt$.  The mismatch parameter
$\Lrod^{2}|k_{\nhat}|/\Drod$ that replaces $\Lrod^{2}/(\Rsph\Drod)$ then
depends on the tilt itself, so on such a host the anchoring angle should be
able to lower the geometric floor and not merely the width of the depleted
region, by turning the rods toward the direction of smaller normal
curvature.  That the two controls separate on the sphere and need not
separate off it is a test of the reading given here, and the case we would
examine next.}

\aud{In summary, the locked tilt controls the polar marginal of hard
spherocylinders on a sphere in two distinct ways.  It sets the width of the
depleted region \mil{that packing induces around each topologically
required singularity}, which is fixed at
roughly six tenths of a rod length under meridian locking and contracts by
nearly an order of magnitude as the director is turned, saturating near
$38^{\circ}$ \pit{on the binned estimator} and so well before latitude locking.  It does not set how
uniform the coverage can be made: the smallest polar-marginal variance
reachable among the sampled tilts follows the geometric combination
$\Lrod^{2}/(\Rsph\Drod)$ with an \har{effective} exponent between $1.1$ and $1.3$ at every
coverage \pit{sampled, most securely below $\eta=0.75$}, so the residual nonuniformity is a property of the rod length,
the host curvature, and the rod diameter rather than of the anchoring
angle.  Whether a director field of the required tilt can be imprinted on a
real shell, for instance magnetically, is the natural experimental question
these results pose\pit{, though we advance it as a prospect rather than a
demonstrated route}.}
\begin{acknowledgments}
We thank Milan Zlatanovi\'{c} (Recognition Physics
Institute) for valuable comments on the manuscript.
\end{acknowledgments}

\section*{AUTHOR DECLARATIONS}
\subsection*{Conflict of Interest}
The authors have no conflicts to disclose.
\subsection*{Author Contributions}
\textbf{Jonathan Washburn}: Conceptualization (equal);
Methodology (equal); Formal analysis (supporting); Writing --
review \& editing (equal).
\textbf{Hartmut L\"owen}: Conceptualization (equal); Formal
analysis (supporting); Writing -- review \& editing (equal).
\textbf{Elshad Allahyarov}: Conceptualization (equal);
Investigation (lead); Methodology (equal); Software (lead);
Formal analysis (lead); Visualization (lead); Writing -- original
draft (lead); Writing -- review \& editing (equal).

\section*{DATA AVAILABILITY}
The data that support the findings of this study are available from
the corresponding author upon reasonable request.

\hypersetup{urlcolor=blue!60!black}

\end{document}